# Telecom-band optical parametric oscillation in a passive silicon microring resonator at cryogenic temperatures

**Shoichiro Yasui[1,a)], Xuejun Xu[1], Hidetaka Nishi[2,3], Haruki Sanada[1], Hiroki Takesue[1]**
[1]Basic Research Laboratories, NTT, Inc., Atsugi, Kanagawa 243−0198, Japan
[2]Device Technology Laboratories, NTT, Inc., Atsugi, Kanagawa 243-0198, Japan
[3]NTT Nanophotonics Center, NTT, Inc., Atsugi, Kanagawa 243-0198, Japan

a) Corresponding author: shouichiro.yasui@ntt.com

**ABSTRACT**

At telecommunications wavelengths, free carriers generated by two-photon absorption limit optical parametric oscillation (OPO) in silicon. Here, we demonstrate telecom-band OPO in a passive silicon microring resonator without electrical carrier sweep-out, driven by a single continuous-wave (CW) pump at 4.5 K. We achieve an on-chip OPO onset power of 2.25 mW and a total conversion efficiency of 1.9% into all parametric sidebands at a pump power of 4.8 mW. Despite CW pumping, the parametric output emerges as a train of pulses synchronized with microsecond-scale self-pulsing. At higher pump powers, OPO emission is absent during intervals of sustained sub-GHz oscillations in the pump transmission, narrowing the temporal window for OPO. Furthermore, by adjusting the pump detuning, we observe OPO at temperatures up to 55 K. These results demonstrate that nonlinear cavity dynamics govern the temporal structure and operating conditions of cryogenic silicon OPOs, providing new insights into controlling the oscillation regimes and temporal characteristics of on-chip nonlinear light sources compatible with cryogenic quantum photonic circuits.

Benefiting from high refractive-index contrast and mature nanofabrication technologies, silicon (Si) photonics has emerged as a platform for dense photonic integration,[1] with applications spanning optical communications,[2] nonlinear light sources,[3] optical sensing,[4] and quantum photonics.[5] The large third-order optical nonlinearity of Si, combined with resonant field enhancement in microring resonators (MRRs), enables efficient four-wave mixing (FWM) at low input powers.[6] Accordingly, Si-MRRs have been investigated as integrated wavelength converters[7] and sources of entangled photon pairs.[8] More recently, cryogenic operation has also been demonstrated in Si quantum photonic circuits that integrate photon-pair sources, optical interferometers, and superconducting nanowire single-photon detectors on silicon-on-insulator substrates.[9]

The nonlinear optical capabilities of Si-MRRs have also been extended to optical parametric oscillation (OPO). To achieve OPO, the parametric gain provided by FWM must exceed cavity losses. In the mid-infrared region, where two-photon absorption (TPA) is relatively weak, frequency-comb generation has been demonstrated using Si-MRRs.[10] At telecommunications wavelengths, however, TPA becomes significant,[11] and the resulting photogenerated free carriers induce nonlinear loss and resonance-frequency shifts through free-carrier absorption (FCA) and free-carrier dispersion (FCD), respectively.[12] Therefore, achieving sufficient FWM gain through a high Q factor and a small mode volume, while suppressing carrier accumulation by shortening the free-carrier lifetime, is crucial for realizing telecom-band OPO in Si.[13]

To address this challenge, reverse biasing of p-i-n junctions integrated into Si-MRRs has been employed to suppress the accumulation of photogenerated free carriers.[14] More recently, soliton-crystal Kerr comb generation at telecom wavelengths has been demonstrated by combining this approach with dispersion engineering.[15] Furthermore, dark-soliton Kerr comb generation in the normal-dispersion regime has been demonstrated using dual-tone pumping and reverse-biased p-i-n junctions.[16] However, these approaches require careful balancing of carrier sweep-out efficiency and doping-induced optical losses, as well as additional junction fabrication and electrical contacts for external biasing.

Cryogenic operation offers an alternative approach to suppressing free-carrier accumulation without introducing carrier-sweep-out junctions or their associated bias wiring. Cooling Si-MRRs has been reported to reduce the free-carrier lifetime from approximately 1.9 ns at room temperature to less than 100 ps below 10 K,[17] potentially mitigating FCA and FCD. Combining the reduced carrier lifetime with resonant field enhancement in a high-Q Si-MRR with a small mode volume therefore offers a promising route to achieving the net parametric gain required for OPO.[13,18] Such an OPO could also serve as an integrated multiwavelength light source for cryogenic silicon quantum photonic circuits.

Nevertheless, to the best of our knowledge, telecom-band OPO in passive Si-MRRs at cryogenic temperatures has not yet been demonstrated.

Here, we demonstrate telecom-band OPO driven by a single continuous-wave (CW) pump in a passive high-Q Si-MRR operated at cryogenic temperatures without electrical carrier sweep-out. Time-resolved measurements further reveal that, despite CW pumping, the OPO output emerges as a periodic train of pulses synchronized with self-pulsing. We also characterize the temperature and pump-detuning dependence of the oscillation characteristics and observe OPO at temperatures up to 55 K under appropriately chosen detuning conditions. These results establish cryogenic operation as an effective approach to realizing telecom-band OPO in passive Si-MRRs and reveal the important role of nonlinear cavity dynamics in determining the OPO operating regime and temporal structure.

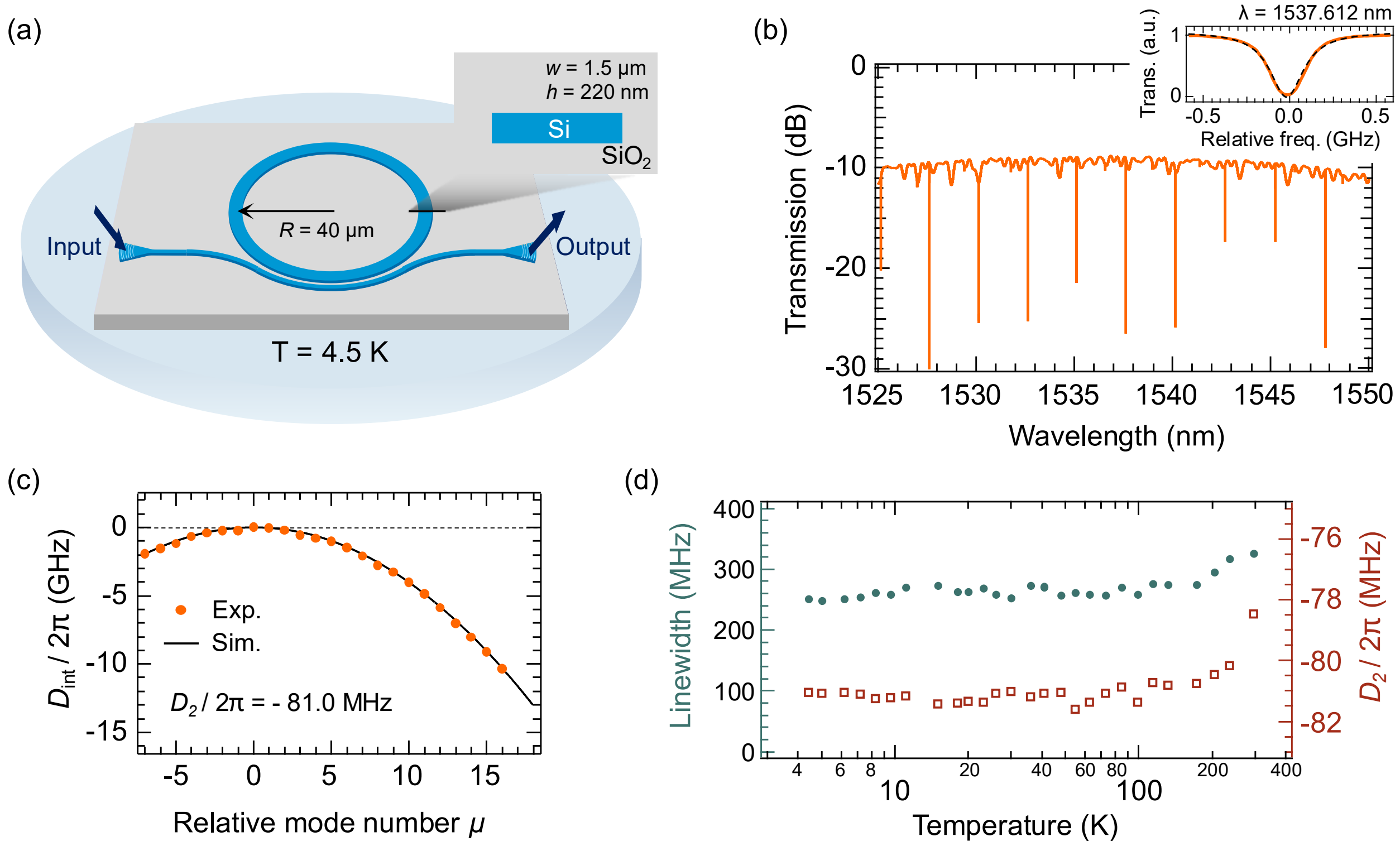

**Fig. 1.** (a) Schematic of the Si-MRR structure and cryogenic measurement setup at 4.5 K. (b) Transmission spectrum measured at 4.5 K. The inset shows the transmission resonance centered at 1537.612 nm. The black dashed line represents a Lorentzian fit. (c) Dispersion of the Si-MRR as a function of mode number. The orange symbols represent experimentally determined values extracted from the measured resonance frequencies, while the black solid line shows numerical results obtained using the finite-element method. (d) Temperature dependence of the resonance linewidth (full width at half maximum, FWHM; left axis, filled teal circles) and second-order dispersion (right axis, open brown squares), extracted from fitting.

To achieve sufficient parametric gain through FWM enhanced by optical confinement in an MRR, it is essential to combine a high Q factor, which provides a long photon lifetime, with a small mode volume, which enables high intracavity optical intensity.[18] However, reducing the resonator size can increase bending radiation and scattering losses, leading to a design trade-off between a high Q factor and a small mode volume.[19,20] To balance these requirements, we employed a high-Q Si-MRR (radius: 40 µm; waveguide width: 1.5 µm; Si thickness: 220 nm) with the same design as that previously characterized as a high-brightness spontaneous four-wave mixing (SFWM) photon-pair source,[21] as shown in Fig. 1(a). Because the ring waveguide supports multiple transverse modes, the directional coupler was designed to selectively excite the target guided mode. The device was mounted in a helium-flow cryostat and cooled to a minimum temperature of 4.5 K. Optical input and output coupling were achieved using grating couplers (GCs), with a coupling efficiency of 36% per port at 4.5 K.

From the transmission spectrum measured at 4.5 K, the pump resonance at 1537.612 nm [Fig. 1(b)] exhibited a loaded Q factor of approximately 7.8 × $10^5$ and a resonance linewidth $\Delta f$ (FWHM) of approximately 250 MHz. The cavity photon lifetime, $\tau_{\mathrm{ph}} = 1/(2\pi\Delta f)$, was estimated to be 0.63 ns. Further details of the experimental setup are provided in the Supplementary Material. The cavity dispersion was characterized by the integrated dispersion $D_{\mathrm{int}}(\mu)$, calculated from the resonance angular frequencies $\omega_\mu$ of the individual cavity modes:

$$D_{\mathrm{int}}(\mu) = \omega_\mu - (\omega_0 + D_1\mu) \approx \frac{D_2}{2}\mu^2$$

Here, μ is the mode number relative to the pumped mode ($\mu = 0$), and $|D_1|/2\pi$ is the free spectral range (FSR) near the pump resonance. From the measured resonances shown in Fig. 1(c), we obtained $D_1/2\pi = -319$ GHz and $D_2/2\pi = -81.0$ MHz. The resonator exhibits normal dispersion near the pump wavelength. For the nearest-neighbor signal–idler mode pair ($\mu = \pm 1$), the cold-cavity frequency mismatch is $(\omega_{+1} + \omega_{-1} - 2\omega_0)/2\pi \approx D_2/2\pi$. Its magnitude (81.0 MHz) is smaller than the resonance linewidth ($\Delta f \sim 250$ MHz), indicating that the dispersion-induced mismatch lies within the resonance linewidth. Figure 1(d) shows the temperature dependence of the resonance linewidth and $D_2/2\pi$. Upon cooling from room temperature to approximately 100 K, the resonance linewidth decreased slightly, while $D_2/2\pi$ shifted toward more negative values, owing to changes in the refractive indices of Si and $SiO_2$. Below 100 K, neither the resonance linewidth nor $D_2/2\pi$ exhibited significant changes.

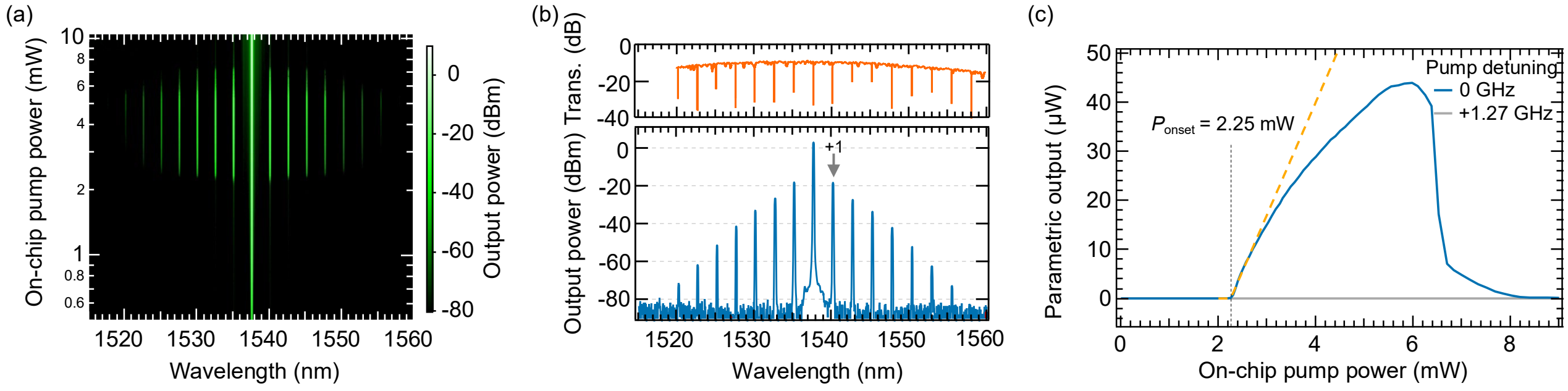

**Fig. 2.** (a) Output spectra measured using an optical spectrum analyzer (OSA) as a function of on-chip pump power. (b) Top: Transmission spectrum of the Si-MRR measured at low input power. Bottom: Representative parametric output spectrum at an on-chip pump power of 4.8 mW. The arrows indicate the pump mode and the first-order (+1) parametric mode used for the analysis. (c) On-chip output power of the +1 parametric mode as a function of on-chip pump power. The blue and gray curves represent measurements at pump detunings of $\Delta f_{\mathrm{p}} = 0$ and +1.27 GHz, respectively. The orange dashed line represents a linear fit to the data in the pump-power range of 2.24–2.74 mW near the OPO onset.

Based on the low-power transmission measurements, the cold-cavity resonance wavelength of the pump mode was determined to be $\lambda_{\mathrm{c}} = 1537.612$ nm. Hereafter, we define the wavelength detuning as $\Delta\lambda_{\mathrm{p}} = \lambda_{\mathrm{p}} - \lambda_{\mathrm{c}}$, where $\lambda_{\mathrm{p}}$ is the pump wavelength, and the frequency detuning as $\Delta f_{\mathrm{p}} = f_{\mathrm{p}} - f_{\mathrm{c}}$, where $f_{\mathrm{c}}$ and $f_{\mathrm{p}}$ are the cold-cavity resonance and pump frequencies, respectively. The on-chip pump power is defined as the power in the bus waveguide after the input GC. We first measured the output spectra at a fixed detuning of $\Delta f_{\mathrm{p}} = 0$ [Fig. 2(a)]. Above approximately 2.2 mW, a sharp increase in sideband power was observed, starting from the cavity modes closest to the pump. In the representative output spectrum at an on-chip pump power of 4.8 mW [Fig. 2(b)], seven sideband pairs were observed. The center wavelengths of the spectral lines closely matched the cavity resonance wavelengths obtained from the low-power transmission measurements, and the sideband pairs were consistent with the energy conservation condition for FWM, $2f_{\mathrm{p}} = f_{\mathrm{s}} + f_{\mathrm{i}}$. Hereafter, we refer to the component generated in the adjacent cavity mode on the longer-wavelength side of the pump (1540.135 nm) as the +1-order parametric output. Figure 2(c) shows the on-chip pump-power dependence of the +1-order parametric output. The output increased sharply at $\Delta f_{\mathrm{p}} = 0$, whereas no comparable increase occurred at $\Delta f_p = +1.27$ GHz. These observations strongly support the interpretation of the observed sideband generation as parametric oscillation driven by cavity-enhanced FWM.

We defined the OPO onset power, $P_{onset}$, as the pump power at which a linear fit to the parametric output immediately above the onset extrapolates to zero. From this analysis, we obtained $P_{onset} = 2.25$ mW. We further defined the on-chip conversion efficiency into all parametric sidebands, excluding the pump mode, as

$$\eta_{tot} = \frac{\sum_{\mu \neq 0} P_\mu}{P_{pump}}$$

where $P_\mu$ is the on-chip output power of each parametric sideband and $P_{pump}$ is the on-chip input pump power. After correcting for the wavelength dependence of the GC losses, we obtained $\eta_{tot} = 1.9\%$ at an on-chip pump power of 4.8 mW, corresponding to a total parametric output power of approximately 93 µW. With increasing pump power, the parametric output did not increase monotonically but reached a maximum near 6 mW before decreasing sharply. This behavior suggests that nonlinear effects in Si modify the cavity resonance conditions and losses at high pump powers.

As the intracavity optical intensity increases, the Kerr effect, TPA, free-carrier effects, and the thermo-optic (TO) effect collectively modify the cavity resonance conditions and losses.[22] The Kerr effect and absorption-induced TO effect red-shift the cavity resonance,[23,24] whereas free carriers generated by TPA induce a blue shift through FCD and additional losses through FCA.[25] The resulting resonance wavelength shift, $\Delta\lambda_{res} = \lambda_{res} - \lambda_c$, can be expressed as

$$\Delta\lambda_{res} \approx \Delta\lambda_{Kerr} + \Delta\lambda_{TO} + \Delta\lambda_{FCD}$$

$$\Delta\lambda_{Kerr} > 0, \qquad \Delta\lambda_{TO} > 0, \qquad \Delta\lambda_{FCD} < 0$$

where the Kerr and TO contributions are positive, while the FCD contribution is negative.[17] While the Kerr response is nearly instantaneous, the free-carrier and thermal responses occur on different timescales. The interplay of these nonlinear effects can destabilize the cavity steady state and give rise to self-pulsing.[26]

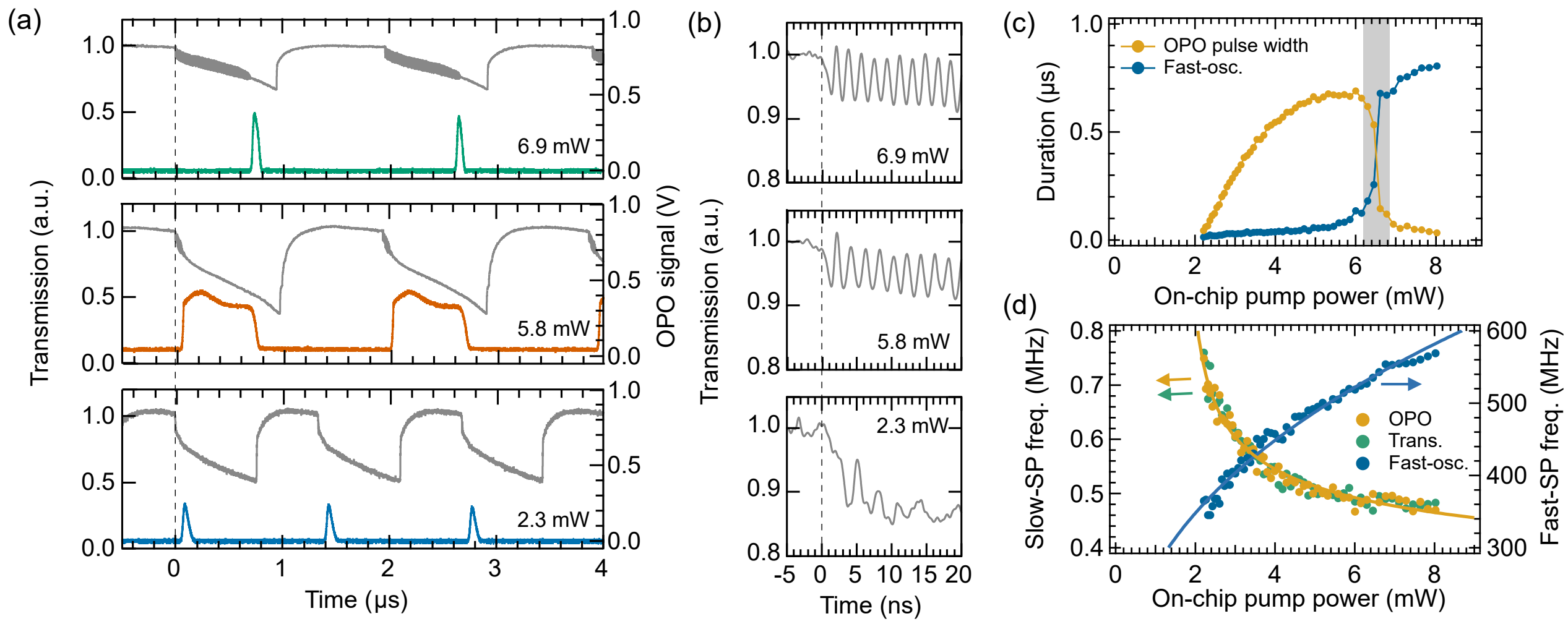

**Fig. 3.** (a) Temporal waveforms of the transmitted pump (gray) and the +1-order OPO output at on-chip pump powers of 2.3, 5.8, and 6.9 mW. Each waveform was acquired by triggering on the sharp falling edge of the transmitted pump signal. (b) Magnified view of the transmitted pump waveform near the trigger time. (c) On-chip pump-power dependence of the OPO pulse width (FWHM; orange) and the duration of the fast oscillatory component (blue). The gray-shaded region indicates the pump-power range where each duration changes abruptly. (d) On-chip pump-power dependence of the slow self-pulsing frequency extracted from the OPO output (orange) and the transmitted pump signal (green) (left axis), and the fast self-pulsing frequency extracted from the high-frequency oscillations in the transmitted pump signal (blue) (right axis). The solid lines show the fitting results for the slow and fast dynamics, respectively.

To investigate the relationship between these nonlinear dynamics and the reduction in OPO output at high pump powers, we simultaneously measured the temporal waveforms of the transmitted pump and the +1-order OPO output [Fig. 3(a)]. At all pump powers, the transmitted pump exhibited periodic variations on a microsecond timescale, and the OPO output appeared as a train of pulses synchronized with these variations. These observations demonstrate that OPO in the Si-MRR at 4.5 K is temporally pulsed in synchrony with self-pulsing (SP), despite CW pumping. A closer examination of the intervals during which the transmitted pump intensity decreased revealed rapid oscillations on a nanosecond timescale [Fig. 3(b)]. Hereafter, we refer to the microsecond- and nanosecond-scale oscillations as slow-SP and fast-SP, respectively.

The duration of fast-SP increased with pump power, exhibiting a sharp increase above approximately 6.7 mW [Fig. 3(c)]. In contrast, the OPO pulse width decreased sharply in the same pump-power range. As shown in Fig. 3(a), at 6.9 mW, OPO emission was absent while fast-SP persisted, and the OPO pulse emerged only after the fast-SP oscillations subsided. At 5.8 mW, however, fast-SP persisted for a shorter interval, and OPO emission extended over a relatively longer duration. This temporal anticorrelation suggests that the nonlinear dynamics responsible for fast-SP constrain the temporal window in which OPO can occur. Furthermore, the pump-power dependence of the OPO pulse width in Fig. 3(c) followed a trend similar to that of the time-averaged OPO output in Fig. 2(c). By averaging the OPO signal over one slow-SP cycle using the time-resolved waveforms, we obtained a pump-power dependence in good agreement with that of the time-averaged parametric output shown in Fig. 2(c) (see Supplementary Material for details). Together, these results support the interpretation that the reduction in time-averaged OPO output at high pump powers is partly attributable to the narrowing of the temporal window for OPO emission.

To investigate the origins of the two SP components, we examined the pump-power dependence of their oscillation frequencies [Fig. 3(d)]. The sub-MHz slow-SP is attributed to thermal–free-carrier SP, arising from the competition between the FCD-induced blue shift and the absorption-induced TO red shift of the cavity resonance on different timescales.[26–28] In contrast, fast-SP exhibits oscillation frequencies of approximately 0.3–0.6 GHz, at which the thermal response can be regarded as quasi-static. Theoretical studies have shown that free-carrier oscillation (FCO) can arise in Si-MRRs through feedback between the intracavity optical field and free carriers generated by TPA, mediated by FCD and FCA.[13] Self-sustained oscillations in the GHz range have also been experimentally reported.[29] Fast-SP may therefore arise from FCO driven by the coupled dynamics of the intracavity optical field and free carriers. Based on these interpretations, we analyzed slow-SP using a phenomenological model that approximates the thermal–free-carrier dynamics as a relaxation-and-switching process, with the oscillation frequency given by

$$f_{\mathrm{slow}}(P_{\mathrm{pump}}) = \left[T_0 + \tau_{\mathrm{rec}} \ln\left(1 + \frac{P_{\mathrm{pump}} - P_0}{P_{\mathrm{T}}}\right)\right]^{-1}$$

where $P_0$ is a reference pump power, $T_0$ is the corresponding oscillation period, $\tau_{\mathrm{rec}}$ is an effective recovery time, and $P_{\mathrm{T}}$ is a characteristic power scale. For fast-SP, we employed the oscillation-frequency relation derived by Cazier et al.[30] in the high-power limit dominated by FCD:

$$f_{\mathrm{fast}}(P_{\mathrm{pump}}) \approx \frac{1}{2\pi}\left(C P_{\mathrm{pump}}^{2/5} - \frac{\Delta\omega}{5}\right),$$

where $\Delta\omega$ is an effective pump detuning in angular frequency and $C$ is a coefficient that depends on the cavity parameters. Details of both models are provided in the Supplementary Material. As shown in Fig. 3(d), the slow-SP frequency decreased with increasing pump power, whereas the fast-SP frequency increased. These contrasting pump-power dependences were well described by the respective models, supporting the involvement of distinct nonlinear dynamics. Furthermore, the abrupt increase in the fast-SP duration near 6.7 mW suggests a change in the nonlinear operating regime of the resonator.

We suggest that the suppression of OPO during fast-SP arises from rapid free-carrier-density oscillations associated with FCO.[13] These oscillations modulate the cavity resonance frequency through FCD and the cavity loss through FCA, thereby causing temporal variations in the effective pump detuning, intracavity pump power, and net parametric growth rate. Because OPO buildup requires a finite time that can substantially exceed the cavity photon lifetime at low net growth rates,[31] such rapid variations can inhibit sufficient parametric amplification during fast-SP. Together, these findings demonstrate that nonlinear cavity dynamics on distinct timescales govern the temporal structure of OPO, with slow-SP determining the pulse repetition period and fast-SP constraining the OPO emission window.

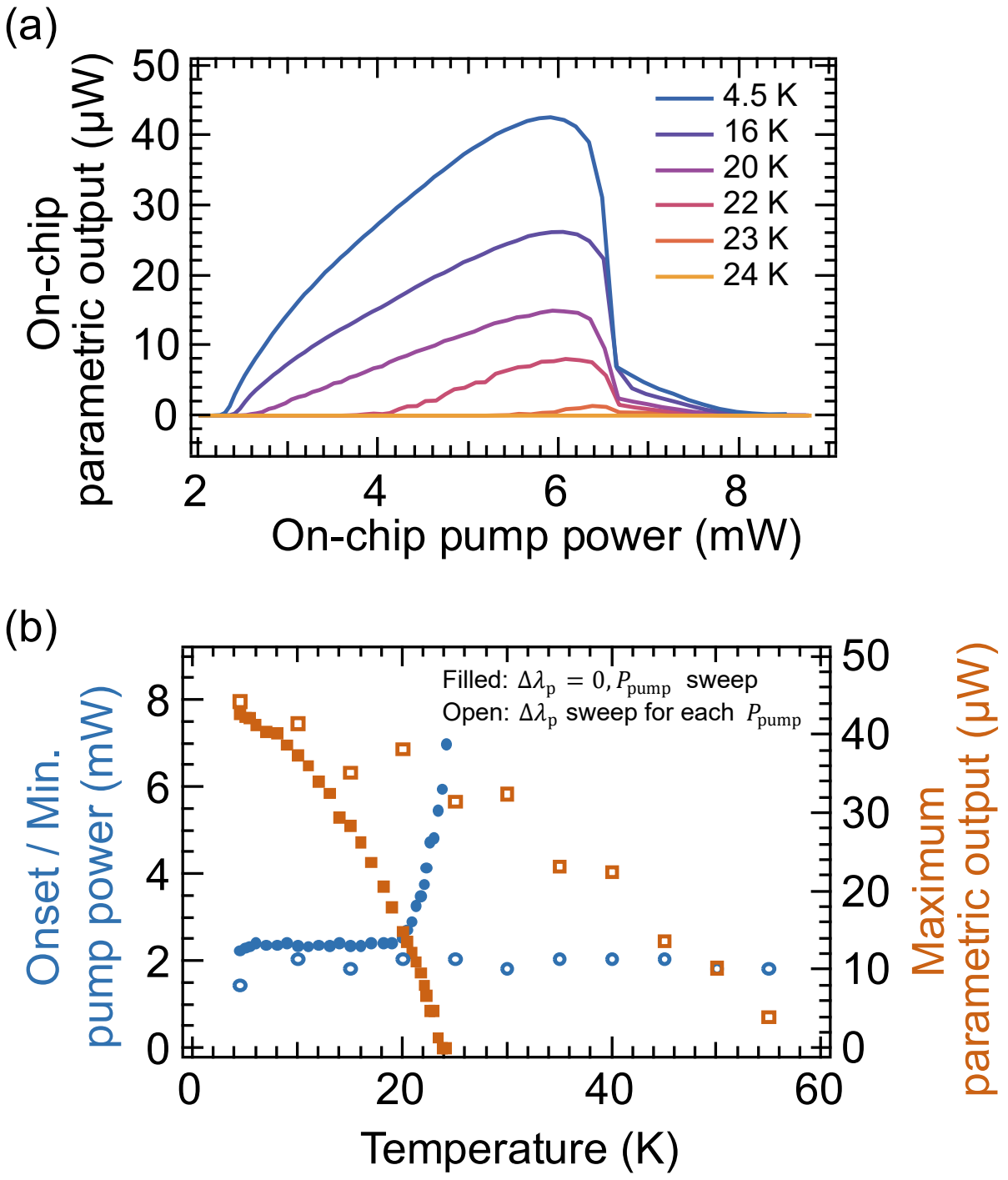

**Fig. 4.** (a) On-chip pump-power dependence of the +1-order parametric output at different temperatures, measured with the pump wavelength fixed at the low-power cold-cavity resonance, $\lambda_c(T)$. (b) Temperature dependence of the OPO onset or minimum pump power (blue circles, left axis) and maximum +1-order parametric output power (orange squares, right axis). Filled and open symbols represent fixed-wavelength power sweeps [panel (a)] and pump-detuning sweeps (see Fig. S3 of the supplementary material), respectively. For the open blue circles, the minimum pump power is defined as the lowest measured power at which the +1-order parametric output exceeds 0.1 μW at any measured detuning.

Finally, we investigated how the OPO operating regime is accessed and its temperature dependence. At each temperature, the pump wavelength was fixed at the cold-cavity resonance, $\lambda_c(T)$, determined from low-power (approximately 1 μW) transmission measurements, and the on-chip pump power was swept [Fig. 4(a)]. The OPO onset power was determined by linearly extrapolating the rising edge of the parametric output, as in Fig. 2(c). For comparison, we also swept the pump detuning at each pump power to determine the minimum pump power at which OPO was observed and the maximum parametric output power obtained under the optimal combination of pump power and detuning [Fig. S3]. The temperature dependences obtained from both measurements are summarized in Fig. 4(b).

Under fixed-wavelength pumping, the OPO onset power increased slightly with temperature up to approximately 20 K but increased sharply at higher temperatures. The maximum parametric output power decreased with increasing temperature, and no output exceeding 0.1 μW was observed above approximately 24 K. In contrast, when the pump detuning was swept, the minimum pump power required to observe OPO remained near 2 mW with only weak temperature dependence. Although the maximum parametric output power decreased with increasing temperature, OPO was observed up to 55 K. At 60 K, no output exceeding 0.1 μW was observed within the investigated range.

As shown in Fig. 1(d), the resonance linewidth and dispersion varied only slightly over the temperature range of 4.5–55 K, consistent with the weak temperature dependence of the lower pump-power boundary of the OPO operating regime observed in the detuning sweeps. In contrast, above 20 K, an enhanced TO response is expected to increase the thermally induced resonance shift, thereby changing the effective pump detuning under fixed-wavelength pumping and increasing the OPO onset power. Indeed, the detuning–power maps [Fig. S3] show that both the thermal triangle and the OPO operating region extend toward positive detuning with increasing temperature, supporting this interpretation. These results demonstrate telecom-band OPO in a passive Si-MRR without electrical free-carrier sweep-out at temperatures up to 55 K through appropriate adjustment of the pump detuning, establishing cryogenic operation as an effective approach to realizing OPO in passive Si-MRRs.

In this work, we demonstrated telecom-band OPO under CW pumping in a passive Si-MRR at 4.5 K, with an on-chip onset power of 2.25 mW. Notably, this milliwatt-scale onset is comparable to the thresholds reported for silicon nitride and silicon carbide MRRs developed for low-threshold OPO,[3,32,33] despite the absence of OPO-specific optimization in our device. This result highlights the potential of Si for low-threshold OPO, benefiting from its large third-order optical nonlinearity and strong optical confinement. At an on-chip pump power of 4.8 mW, we achieved a total conversion efficiency of 1.9%, corresponding to approximately 93 μW of time-averaged parametric output. In contrast to the previously reported picowatt-level sidebands without a clear OPO threshold in a passive Si microresonator at room temperature,[34] our device exhibits a distinct onset and microwatt-level output without electrical carrier sweep-out.

Self-pulsing OPO under CW pumping has previously been observed in bulk resonators through thermal feedback and optical bistability,[35] while Kerr-soliton breathing has been reported in SiN microresonators.[36] In our cryogenic Si-MRR, two distinct nonlinear dynamical timescales shape the OPO output: slow-SP determines the pulse repetition period, whereas OPO emission is absent during fast-SP, with longer fast-SP durations accompanied by narrower OPO emission windows. These dynamics are consistent with thermal–free-carrier and optical-field–free-carrier feedback, respectively, which can modify the temporal balance between parametric gain and cavity loss. Thus, although free carriers generally limit telecom-band OPO in Si, their nonlinear feedback introduces an additional dynamical degree of freedom that could be exploited to control the temporal characteristics of OPO.

We observed OPO at temperatures up to 55 K by appropriately selecting the pump power and detuning. The suppression of carrier accumulation and FCA resulting from the shortened free-carrier lifetime at cryogenic temperatures is considered an important factor enabling OPO.[13,17] Meanwhile, as the OPO operating region changes with temperature (Fig. S3), the scan path also plays an important role in accessing the nonlinear operating states that support oscillation. A quantitative understanding of these operating conditions and nonstationary dynamics requires device- and setup-specific characterization of the free-carrier lifetime and thermal relaxation parameters, together with modeling based on an extended Lugiato–Lefever equation incorporating free-carrier and thermal dynamics.[22,37] In particular, the photon lifetime of our resonator (approximately 0.63 ns) is longer than the free-carrier lifetime of less than 100 ps reported for Si below 10 K,[17] making it difficult to independently determine carrier relaxation from the temporal response of the cavity transmission alone. Independent characterization using nonresonant waveguides or other methods would enable a more quantitative understanding of the operating conditions for OPO, slow-SP, and fast-SP.

Previous demonstrations of cryogenic integrated quantum photonic circuits[9] and electro-optic operation of Si microresonators at cryogenic temperatures[38,39] motivate the development of integrated optical frequency comb sources for cryogenic photonic systems. With further dispersion and mode engineering,[10,40] together with control of self-pulsing dynamics, passive cryogenic Si-MRR OPOs could be extended toward low-noise Kerr-comb states, including soliton crystals[15] and dark-pulse combs.[16] Such sources could serve as versatile nonlinear photonic building blocks for integrated quantum photonic circuits.

In this work, we demonstrated telecom-band OPO driven by a single CW pump in a passive Si-MRR without electrical free-carrier sweep-out. At 4.5 K, we achieved an on-chip OPO onset power of 2.25 mW and an on-chip conversion efficiency of 1.9% into all parametric sidebands at a pump power of 4.8 mW. Time-resolved measurements revealed that, despite CW pumping, the OPO output emerged as a periodic train of pulses synchronized with microsecond-scale SP. At higher pump powers, little OPO emission was observed during intervals of sustained sub-GHz oscillations in the transmitted pump, suggesting that the narrowing of the OPO emission window contributed to the reduction in time-averaged output power. Furthermore, although the resonance linewidth and dispersion exhibited relatively weak temperature dependences, the OPO operating regime depended strongly on temperature and pump detuning, with oscillation observed up to 55 K through appropriate detuning control. These results demonstrate that the operation of cryogenic Si OPOs is strongly influenced not only by static cavity properties but also by nonlinear cavity dynamics, providing new insights into the design of operating conditions and control of temporal characteristics of integrated nonlinear light sources for cryogenic quantum photonic circuits.

## SUPPLEMENTARY MATERIAL

See the supplementary material for details of the experimental setup and calibration procedures, models and fitting procedures for slow- and fast-SP, cycle-averaged OPO signals derived from time-resolved measurements, and temperature-dependent pump-detuning–power maps.

## ACKNOWLEDGMENTS

This work was supported by JSPS KAKENHI Grant Number JP23K26580.

## AUTHOR DECLARATIONS

### Conflict of Interest

The authors have no conflicts to disclose.

### Author Contributions

**Shoichiro Yasui**: Conceptualization (lead); Methodology (lead); Investigation (lead); Formal analysis (lead); Data curation (lead); Visualization (lead); Writing – original draft (lead); Writing – review & editing (lead). **Xuejun Xu**: Resources (supporting); Writing – review & editing (supporting). **Hidetaka Nishi**: Resources (supporting); Writing – review & editing (supporting). **Haruki Sanada**: Writing – review & editing (supporting). **Hiroki Takesue**: Supervision (lead); Project administration (lead); Writing – review & editing (supporting).

## DATA AVAILABILITY

The data that support the findings of this study are available from the corresponding author upon reasonable request.